\documentclass[preprint,landscape]{aastex631}

\usepackage{amsmath} 

\usepackage{bm} 

\usepackage[colorinlistoftodos,textsize=footnotesize]{todonotes}
\definecolor{kellygreen}{rgb}{0.3, 0.73, 0.09} 
\definecolor{palegreen}{rgb}{0.6, 0.98, 0.6}
\definecolor{hlgreen}{rgb}{0.1, .6, 0.05} 

\usepackage{soul}  

\newcommand{\cnote}[1]{\todo[inline,backgroundcolor=palegreen!25,bordercolor=gray]{#1}}
\renewcommand{\cnote}[1]{\relax}

\newcommand{\inote}[1]{\textcolor{hlgreen}{\textbf{\textsl{[[#1]]}}}}
\renewcommand{\inote}[1]{\relax}

\newcommand{\bC}{\mathbf{C}}

\newcommand{\bx}{\mathbf{x}}

\newcommand{\btheta}{\bm{\theta}}

\newcommand{\bSigma}{\bm{\Sigma}}

\newcommand{\ngal}{678,239}  

\newcommand{\pz}{photo-$z$}

\renewcommand{\ll}{\ell} 

\newcommand{\wv}{\lambda}  
\newcommand{\lw}{\ell}  
\newcommand{\fden}{F}  
\newcommand{\cb}{\phi}  
\newcommand{\lp}{\kappa}  
\newcommand{\gb}{\varphi}  
\newcommand{\vgb}{\bm{\varphi}}  

\newcommand{\fest}{f}  
\newcommand{\vfest}{\vec{f}}  
\newcommand{\prcn}{\tau}  
\newcommand{\vfden}{\vec{F}}  
\newcommand{\dmat}{\bm{X}}  
\newcommand{\pmat}{\bm{T}}  

\newcommand{\coef}{\theta}  
\newcommand{\vcoef}{\bm{\theta}}  
\newcommand{\vcmcoef}{\bm{m}}  
\newcommand{\lchar}{\chi}  
\newcommand{\vmcoef}{\bm{\mu}}  
\newcommand{\ccoef}{\bm{\Sigma}}  

\newcommand{\iw}{k}  
\newcommand{\nw}{K}  
\newcommand{\io}{o}  
\newcommand{\no}{O}
\newcommand{\ib}{b}  
\newcommand{\ic}{c}  
\newcommand{\nc}{C}
\newcommand{\il}{l}  
\newcommand{\nl}{L}
\newcommand{\nb}{B}  

\newcommand{\like}{\mathcal{L}}  
\newcommand{\mlike}{\mathcal{M}}  

\DeclareMathOperator{\N}{N}

\DeclareMathOperator{\Corr}{Corr}
\DeclareMathOperator*{\argmax}{arg\,max}
\DeclareMathOperator{\tr}{Tr}

\newcommand{\dif}{\textrm{d}}  

\newcommand{\Norm}{\mathop{\mathrm{Norm}}}

\DeclareMathOperator{\expect}{\mathbb{E}}

\DeclareMathOperator{\cov}{\mathrm{Cov}}

\let\swpack=\texttt
\newcommand{\stan}{\swpack{Stan}}
\newcommand{\pymc}{\swpack{PyMC}}

\begin{document}


\title{Splines 'n Lines:  Rest-frame galaxy spectral energy distributions\\
via Bayesian functional data analysis}

\author[0000-0003-2880-1216]{David Kent}
\affiliation{Department of Statistics and Data Science \\
Cornell University \\
Comstock Hall \\
Ithaca, NY 14853, USA}

\author[0000-0002-7034-4621]{Tam\'as Budav\'ari}
\affiliation{Department of Applied Mathematics \\
Whitehead Hall \\
Johns Hopkins University \\
Baltimore, MD 21218, USA}

\author[0000-0003-4692-4607]{Thomas J. Loredo}
\affiliation{Cornell Center for Astrophysics and Planetary Science \\
Cornell University \\
Space Sciences Building \\
Ithaca, NY 14853, USA}
\affiliation{Department of Statistics and Data Science \\
Cornell University \\
Comstock Hall \\
Ithaca, NY 14853, USA}

\author[0000-0002-6713-2257]{David Ruppert}
\affiliation{Department of Statistics and Data Science \\
Cornell University \\
Comstock Hall \\
Ithaca, NY 14853, USA}
\affiliation{School of Operations Research and Information Engineering \\
Cornell University \\
Rhodes Hall \\
Ithaca, NY 14853, USA}


\correspondingauthor{Thomas J. Loredo}

\begin{abstract}
Survey-based measurements of the spectral energy distributions (SEDs) of galaxies have flux density estimates on badly misaligned grids in rest-frame wavelength.
The shift to rest frame wavelength also causes estimated SEDs to have differing support.
For many galaxies, there are sizeable wavelength regions with missing data.
Finally, dim galaxies dominate typical samples and have noisy SED measurements, many near the limiting signal-to-noise level of the survey.
These limitations of SED measurements shifted to the rest frame complicate downstream analysis tasks, particularly tasks requiring computation of functionals (e.g., weighted integrals) of the SEDs, such as synthetic photometry, quantifying SED similarity, and using SED measurements for photometric redshift estimation.
We describe a hierarchical Bayesian framework, drawing on tools from functional data analysis, that models SEDs as a random superposition of smooth continuum basis functions (B-splines) and line features, comprising a finite-rank, nonstationary Gaussian process, measured with additive Gaussian noise.
We apply this \emph{Splines 'n Lines} (SnL) model to a collection of 678,239 galaxy SED measurements comprising the Main Galaxy Sample from the Sloan Digital Sky Survey, Data Release 17, demonstrating capability to provide continuous estimated SEDs that reliably denoise, interpolate, and extrapolate, with quantified uncertainty, including the ability to predict line features where there is missing data by leveraging correlations between line features and the entire continuum.
\inote{We could add about 50 words here.}
\end{abstract}

\section{Introduction}\label{intro}

Until the advent of astronomical spectroscopy, astronomy was concerned solely with the study of the motion and brightness of stars.
Knowledge about the physical nature of stars was deemed inaccessible.
French philosopher Auguste Comte wrote in 1835, ``We understand the possibility of determining their [celestial bodies'] shapes, their distances, their sizes and their movements; whereas we would never know how to study by any means their chemical composition, or their mineralogical structure, and, even more so, the nature of any organized beings that might live on their surface... [E]very notion of the true
mean temperatures of the stars will necessarily always be concealed from us'' \citep{H10-SpecHist}.
Johann Z\"ollner, an astronomer at Leipzig University (who would later help pioneer astrophysics and confirm Christian Doppler's theory that motion alters the spectrum of stars), in response to a physicist colleague's question about the nature of the stars, asserted, ``What the stars are, we do not know and will never know!'' \citep{HK84-HistAstroH2H}.

Around the time of these assertions in the early 1800s, physicists were already developing the tools that would prove them wrong: the tools of spectroscopy.
It was known since {}Newton's experiments with prisms ca.\ 1666 that white sunlight is a mixture of light of many colors; Newton coined the term \emph{spectrum} for the band of smoothly dispersed colors formed by a prism.
But it was not until the early 1800s that scientists began to make precise measurements of spectra.
Herschel measured the distribution of heat in the Sun's spectrum with a thermometer, and found that the temperature was maximized beyond the red end of the spectrum---the discovery of infrared radiation.
Johann Ritter used silver chloride, which darkens on exposure to light, to study the Sun's spectrum, and found darkening beyond the violet end---the discovery of ultraviolet radiation.
The most impactful discoveries concerned observations of \emph{spectral lines}---bands of absorption and emission in an otherwise smooth spectrum.
Joseph von Fraunhofer invented the first spectroscope or spectrometer, capable of measuring the locations of features in spectra.
He observed bright lines in the spectra of flames---emission lines.
Turning his spectrometer to the Sun, he found dark absorption lines at the same locations of the emission lines seen in flames colored with different chemicals.
Spectroscopy soon enabled the measurement of the temperatures of stars and the compositions of stellar atmospheres.
It is widely credited for the birth of astrophysics, leading to a detailed understanding of the physics of stars and other celestial bodies \citep{H14-SpecHist}.

In 1899, Scheiner reported the first recorded spectrum of a \emph{galaxy}, M31; visual analysis of the spectrum indicated that M31 was an assembly of stars rather than a cloud of gas (see \citealt{R95-SpecHist} for a historical survey of galaxy spectroscopy).
Within just 30 years, Lema\^itre and Hubble used spectra of a few dozen galaxies (mostly observed by Slipher), along with brightness-based distance estimates, to argue that the universe is expanding, with galaxies receding from each other with velocities proportional to distance---the Hubble-Lema\^itre law \citep{R95-SpecHist,L11-Lemaitre}.
Spectroscopy thus quickly proved as transformational for extragalactic astronomy and cosmology as it had become for stellar astrophysics.


The Sloan Digital Sky Survey (SDSS)---the first large-scale automated digital sky survey---has vastly expanded the scope of galaxy spectroscopy by producing large catalogs of galaxy spectra.
It's Main Galaxy Sample (MGS; \citet{S+02-SDSS-MGS}) comprises nearly 700,000 galaxies, enabling detailed study of galaxy spectra at the population level.

A notable feature of SDSS spectra is that they are \emph{spectrophotometrically calibrated}; they can be used, not only to measure the local shape of star and galaxy spectra and the locations of lines, but also to measure absolute flux, including across broad regions in wavelength (see \citealt{SDSS-DR6} and \citealt{Y+16-SDSSSpecPhoto} for discussions of SDSS spectrophometric calibration).
Put differently, a spectrophotometrically calibrated spectrum provides a faithful estimate of the \emph{spectral energy distribution} (SED) of an object, which we denote by $F(\lambda)$, the energy flux density per unit wavelength (and per unit time and area).
Mathematically, spectrophotometric calibration implies that \emph{functionals} of measured spectra---mappings from the SED function to a scalar, typically via integration---are meaningful and accurate (see \citealt{W+20-SpecPhotoFunc} for a discussion of spectrophotometric calibration from a functional analysis perspective).
A common example is computing the flux in a photometric band (found by integrating the product of the SED and the photometric response function associated with the band's filter).
Another example is computing pairwise similarities between SEDs in order to discover structure in the population of SEDS, e.g., using manifold learning techniques (e.g., \citealt{LBM16-SDSSSpecManifold}).

The work we report here is motivated by photometric redshift estimation (\pz).
Generative (forward-modeling) \pz\ approaches use some kind of SED model (most simply, a set of ``template'' SEDs for prototypical galaxies) and synthetic photometry to predict color as a function of redshift and galaxy type.
Comparing predicted and observed colors (and potentially magnitudes) then enables estimation of redshift (and type).
At a minimum, such methods need to do accurate synthetic photometry over a template library.
For more sophisticated methods, other functionals of the SEDs play a role.
A new method our team is developing (to be described elsewhere) uses the entire SDSS MGS catalog (rather than a few prototype SEDs) to build a low-dimensional continuous SED model for \pz.
It requires computing pairwise rest-frame SED similarities for all measured SEDs.

\begin{figure}
\includegraphics[width=\linewidth]{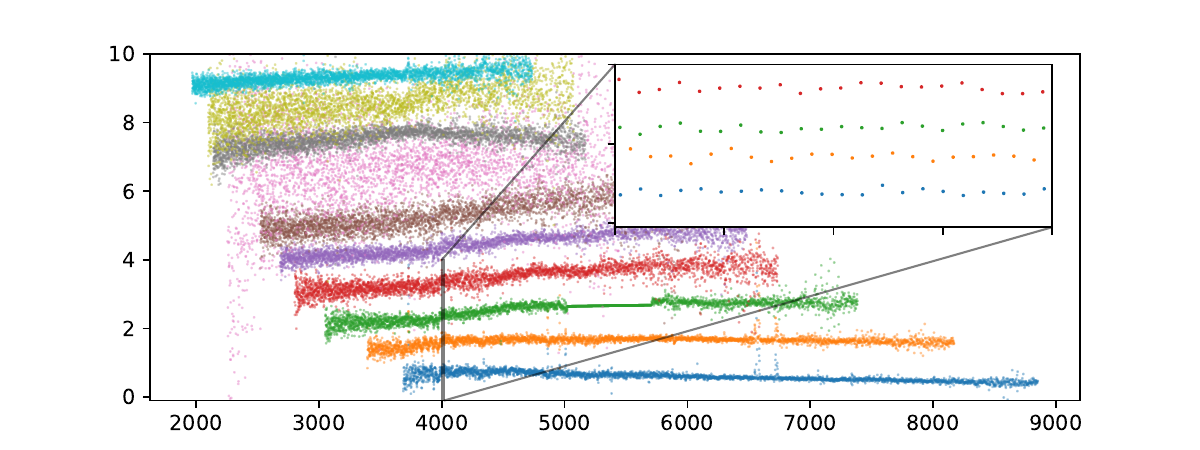}
\caption{A sample of measurements of 10 galaxy SEDs as a function of rest-frame wavelength, chosen from across the range of redshifts in the SDSS MGS, plotted as small points (colors distinguish the 10 galaxies; measured SEDs are offset vertically for visual separation).
Missing intervals---flagged by flux table entries with precision (inverse variance) equal to 0---are linearly interpolated by convention; this can be seen in the third dataset from the bottom (green points).
Inset zooms in on a small wavelength range, showing that the measurements are not aligned in wavelength.}
\label{example_seds}
\end{figure}

A significant challenge in exploiting spectrophotometric SEDs is that the measurements for different sources are typically not aligned in wavelength.
Instrumental drifts and barycentric corrections contribute to this.
But the problem is particularly severe for galaxy SEDs, where the astrophysically fundamental quantity is the \emph{rest-frame SED}.
Since every galaxy has an essentially unique redshift, even if galaxy spectra are observed on a single fixed wavelength grid, the grids will be badly misaligned in rest-frame wavelength.
For similar reasons, the measured SEDs for different galaxies will have different support (wavelength span) in the rest frame.
And many measured SEDs have significant gaps due to uncorrectable problems with the data.
Figure~\ref{example_seds} shows 10 measured SEDs from the MGS that illustrate these issues.

\cnote{Note to DK: Please label axes of figures. The 10-SEDs plots could use ``Relative flux'' as the ordinate label.}

Galaxy SEDs are not infinitely diverse; many SEDs bear a family resemblence, and SED similarity can be used to identify galaxy classes or types.
This suggests that information could be shared across SEDs to ``fill in the gaps'' and enable interpolation and modest extrapolation of SED measurements.
Also, besides the challenges just desribed, SED measurements are noisy, more so for dim galaxies, which are more prevalent than bright ones (due to both the shape of the galaxy luminosity function, and geometry, with the volume at the distant, faint edge of a survey greatly exceeding the nearby volume).
Resemblance across SEDs offers the potential to ``denoise'' by some kind of smoothing that exploits the resemblance (e.g., ``shrinkage'' in the parlance of statistics).

\cnote{Add some remarks/pointers to existing poorly/partially-described methods, e.g., using PCA.}

We describe here a probabilistic model for a catalog of observed SEDs that estimates the underlying true SEDs across the catalog in a way that identifies and exploits similarity in SED shapes.
It is an example of Bayesian functional data analysis (FDA).
FDA is a mature area of statistics that models \emph{populations of functions}, rather than the populations of scalars and vectors that are the fundamental ``units'' of more conventional statistical methods (see, e.g., \citealt{ZMW04-FDALongData,RS05-FDABook,RHG09-FDA-R-MATLAB,WCM16-FDAReview}).
Bayesian FDA implements FDA using a hierarchical Bayesian approach, with separate probabilistic layers modeling the population of functions, and modeling measurement errors.

In the next section we describe the main features of the model (technical details are in an Appendix).
In section~\ref{results} we present results from applying the model to the measured SEDs in the SDSS MGS catalog.
We briefly discuss our findings in section~\ref{discuss}.

\section{Model overview}\label{model}

\cnote{Update the catalog numbers; some numbers here and at the start of \S~\ref{results} may be from an older sample that is somewhat larger than MGS. Here, ``over a billion'' was $\approx 3.2\times 10^9$.}

The SDSS MGS SED catalog is large, comprising over a billion flux measurements across over half a million unaligned wavelength grids.
This motivates adopting a SED model that aims to combine simplicity and flexibility, ideally one with closed-form expressions for many quantities of interest.
In addition, the SDSS SED catalog supplements measured spectra with important derived quantities: measurements of the areas and widths of up to 32 well-studied spectral lines with known restframe wavelengths (these measurements are produced by the \texttt{spectro1D} pipeline, \citealt{S+02-spectro1d}).
The line locations are used to estimate the redshift.
Because of the importance of spectral lines for galaxy physics and redshift estimation (including their impact on photometric redshift), the SED model should exploit line fitting information.

\subsection{Linear individual SED model}

Motivated by these considerations, our departure point is a \emph{linear model} for SEDs that models the continuum with a smooth expansion in localized components (flexible enough to accomodate sharp features like breaks and small lines), and separately models the lines that are identified as important in the SDSS pipeline and thus explicitly fit.
Focusing at first on the flux density for a single galaxy, we write it as a linear superposition of continuum and line basis functions:
\begin{equation}
\fden(\lw) = 
  \sum_{\ib=1}^{\nc} \coef_\ib \, \cb_\ib(\lw)
  + \sum_{\ib=\nc+1}^{\nc+\nl} \coef_\ib \, \lp(\lw; \lchar_{\ib-\nc}),
\label{fdmodel-cl}  
\end{equation}
where $\cb_\ic(\lw)$ is a continuum basis function, and $\lp(\lw;\lchar)$ is a unit-area line profile as a function of wavelength and line characteristics $\lchar$ (comprising the known wavelength of the line, and its estimated width).
The somewhat awkward indexing allows us to gather the linear coefficient parameters into a single parameter vector, $\vcoef = (\coef_1,\cdots,\coef_{\nc+\nl})$, comprising all $\nc+\nl$ coefficients $\coef_\ib$; see Table~\ref{tbl:indx} for a description of the indices, and Table~\ref{tbl:symbols} for definitions of the symbols introduced here and below.
We similarly gather the basis functions together as 
\begin{equation}
\gb_\ib(\lw) =
  \begin{cases}
    \cb_\ib(\lw) & \mbox{for } \ib=1 \mbox{ to } \nc \\ 
    \lp(\lw; \lchar_{\ib-\nc}) & \mbox{for } \ib=\nc+1 \mbox{ to } \nc+\nl.
  \end{cases}
\label{gbasis}  
\end{equation}
Denote the total number of coefficients and basis functions (per SED) as $\nb = \nc+\nl$.
Then the flux density model can alternatively be written more simply as
\begin{equation}
\fden(\lw) = 
  \sum_{\ib=1}^{B} \coef_\ib \, \gb_\ib(\lw).
\label{fdmodel-g}  
\end{equation}

We consider the line characteristics, $\{\lchar_\il\}$ for $\il=1$ to $\nl$, to be fixed; in the calculations of \S~\ref{results} we set them equal to the best-fit values reported by SDSS DR17.
Mathematically, this is helpful in that it keeps the model linear; astrophysically, it reflects a focus on the total flux in a line (given by a line's coefficient), with the line width being of secondary importance (the line locations are known once redshift is estimated).

\begin{deluxetable}{ll}[t]\label{tbl:indx}
\tablecaption{Index variables}

\tablehead{\colhead{Symbol} & \colhead{Definition} \\ 
\colhead{} & \colhead{} } 

\startdata
$\io$  & object (galaxy) index, $1\ldots \no$ \\ 
$\iw$  & wavelength index, $1\ldots \nw_\io$ for object $\io$ \\ 
$\ib$  & basis function index (continuum \& lines), $1\ldots \nb$ \\ 
\enddata

\end{deluxetable}

\begin{deluxetable}{ll}[t]\label{tbl:symbols}
\tablecaption{Symbols used for the individual SED and population models}

\tablehead{\colhead{Symbol} & \colhead{Definition} \\ 
\colhead{} & \colhead{} } 

\startdata
$\lambda$  & wavelength \\ 
$\lw$  & base-10 logarithm of wavelength in \AA \\ 
$\fden$  & flux density (per unit wavelength) \\ 
$\fest$  & flux estimate (data product) \\ 
$\prcn$  & flux estimate precision (inverse variance, data product) \\ 
$\cb_\ic(\lw)$  & continuum basis functions (B-splines), $\ic=1$ to $\nc$ \\ 
$\lp(\lw;\lchar)$  & unit-area line profile with characteristics $\lchar$ \\ 
$\gb_\ib(\lw), \vgb(\lw)$  & collected basis functions (splines 'n lines) \\ 
$\coef_\ib, \vcoef$  & basis function coefficients (for continuum \& lines) \\ 
$\vmcoef$  & mean for population distribution for coefficients \\ 
$\ccoef$  & covariance matrix for population distribution for coefficients \\ 
\enddata

\end{deluxetable}

Cataloged spectra are ``1D spectra'' produced by pipelines processing more complex raw data (e.g., 2D images of cross-dispersed light).
Henceforth, ``spectral data'' refers to the processed and calibrated 1D spectra.

Spectral data measure functionals of the continuous SEDs, over pixel grids in (nominal) wavelength space that differ from object to object.
Formally, associated with each pixel is a line spread function, and the measurement is of a functional of the SED---a local average of the SED corresponding to an integral of the SED times the line spread function.
Here we make the common assumption that the line spread function is narrow enough that the functional may be approximated by simply evaluating the SED at a \emph{nominal wavelength} for the pixel, which we henceforth simply call the pixel wavelength.

We denote the pixel log-wavelengths for object $\io$ as $\lw_{\io\iw}$, with $\iw$ indexing the wavelengths, from 1 to $\nw_\no$.
The catalog data reports measurements of $\fden_{\io\iw} \equiv \fden(\lw_{\io\iw})$.

The collection of flux densities for object $\io$ can be written
\begin{align}
\fden_{\io\iw} 
  &= \sum_{\ib=1}^{\nc} \coef_{\io\ib} \, \cb_\ib(\lw_{\io\iw})
     + \sum_{\ib=\nc+1}^{\nc+\nl} \coef_{\io\ib} \, \lp(\lw_{\io\iw}; \lchar_{\io,\ib-\nc} ) \\
  &= \sum_{\ib=1}^{\nb} \coef_{\io\ib} \, \gb_\ib(\lw_{\io\iw})
     \qquad \text{for } \iw=1\text{ to }\nw_\io.
\label{fdens-all}
\end{align}

There are three types of multi-component spaces in our model:
\begin{itemize}
\item The $\nb$-element space of basis functions (considering the continuum and line bases collectively), spanned by the $\ib$ index of the coefficients $\coef_{\io\ib}$;
\item The $\no$-dimensional space of objects (galaxies), spanned by the $\io$ indices;
\item The wavelength grids, with the grid for object $\io$ having $\nw_\io$ elements, spanned by the $\iw$ indices.
\end{itemize}
 We use vector notation to simplify some equations by suppressing the $\ib$ and $\iw$ indices.
 We use bold math to denote vectors suppressing the basis function index, $\ib$, as in $\vcoef_\io$ for the coefficients for object $\io$.
 We use arrows to denote vectors suppressing the wavelength index, $\iw$, as in $\vfden_\io$ for the predicted spectrum for object $\io$, and $\vfest_\io$ for the measured spectrum (described below).
 Note that the latter vectors over wavelengths have different dimension for each object.

\cnote{Currently we're not using the wavelength vectors; perhaps in an appendix?}

\subsection{Hierarchical Bayesian model for the SED population}

If we knew the ``true'' coefficients describing a particular galaxy's SED, we could compute functionals of the SED, either by using the SED model to evaluate the SED on a convenient wavelength grid (e.g., from a quadrature rule), or by using the basis functions to compute functionals without a grid (i.e., using integrals involving the basis functions, which may be computable analytically for some functionals).

Of course, noise in the measurements and missing data (within a SED, or beyond its edges after shifting to the rest frame) mean that there will be uncertainty in the estimated coefficients.
The uncertainty is especially problematic for wavelength regions with gaps.

A hierarchical Bayesian model for a SED population aims to fit the SED coefficients for a large ensemble of SEDs \emph{jointly}, with the goal of sharing information across related SEDs to discover and exploit relationships between coefficients.
The model has two probabilistic components (whence the ``hierarchical'' qualifier): each SED is treated as a random sample from a SED population, whose measurement is then subject to random measurement error.

Hierarchical Bayesian (HB) models have become increasingly popular for demographic modeling in astronomy over the last two decades; see Loredo \citeyearpar{L13-BackwardLook} for an overview of key ideas and a survey of the literature as of 2013.
HB models can serve two purposes.
Most commonly in astronomy, such models are used to learn population distributions for a class of objects, accounting for measurement effects (both measurement error and, when relevant, selection effects).
With population inference as the goal, uncertainties in the parameters for the members of the population are ``nuisance parameters'' and get marginalized over, producing a marginal posterior distribution for the population parameters.
In many applications outside of astronomy, the goal instead is to exploit membership in a population to improve the estimates of properties of the members.
In such settings, it is the population parameters that are nuisance parameters.
When there is significant uncertainty in the learned population parameters, that can be propagated through the member-level inferences via marginalization over the population parameters.
When the data are voluminous so the population parameters are well-estimated, an \emph{empirical Bayes} approximation that optimizes rather than marginalizes over the population parameters is often adequate.

Here we are in the latter situation: the goal of our model is to improve the estimates of the SEDs by sharing information (``borrowing strength'' is the statistics parlance) across measured SEDs.
This enables improved denoising of SEDs, and especially helps with interpolation across gaps, and extrapolation past the rest frame support of a particular SED's data (so long as we restrict extrapolation to regions where there are data from many other SEDs).


\begin{figure}
\centerline{\includegraphics[width=.4\textwidth]{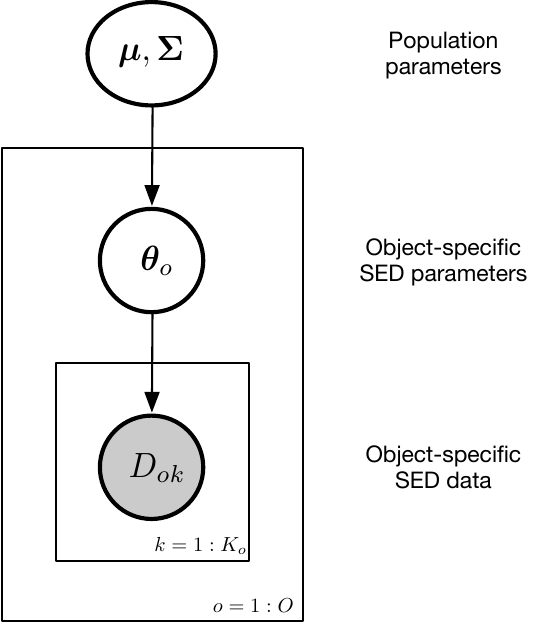}}
\caption{Directed acyclic graph (DAG) for the SED hierarchical Bayesian model.}
\label{SnL-DAG}
\end{figure}

Figure~\ref{SnL-DAG} depicts the structure of our model as a directed acyclic graph (DAG), indicating the dependence relationships among the probabilistic components of the model.
The nodes (round shapes enclosing symbols) represent a priori uncertain quantities treated as random variables.
The top node contains parameters describing the population, here the mean vector and covariance matrix for the population of SED coefficents (described further below).
The middle node contains the coefficient vector describing a particular SED.
The shaded node, here containg $D_{\io\iw}$, denotes the data from object $o$ measuring the flux at wavelength $\lw_\iw$; the gray shading indicates a quantity that becomes known once data are available.
The arrows depict conditional dependence, so each node represents a probability distribution for the node's quantities, conditional on the quantities from any parent nodes.
The plates (square boxes annotated with an index at the lower right) indicate conditionally independent replication of a group of nodes.
Here the outer plate replicates over the objects (galaxies), and the inner plate replicates over wavelengths for a particular object. 

The DAG describes how one can compute a joint distribution for all of the quantities in the graph.
For this graph, we can construct the joint distribution by reading down from the top:
\begin{equation}
p(\vmcoef, \ccoef, \{\vcoef_\io\}, \{D_{\io\iw}\}) =
  p(\vmcoef, \ccoef) \prod_{\io=1}^\no \left[ p(\vcoef_\io|\vmcoef, \ccoef)
    \prod_{\iw=1}^{\nw_\io} p(D_{\io\iw}|\vcoef_\io) \right].
\label{joint1}
\end{equation}
The first factor is a prior distribution for the population parameters.
The product over objects corresponds to the out plate, and the first factor in the product evaluates the population distribution for each SED's coefficients.
The last product over wavelengths for a particular SED accounts for measurement error in the flux estimates.
Since the data are known (in shaded nodes), what matters in the last factor is the dependence of the data probabilities on the SED parameters, $\vcoef_\io$.
That is, we need to specify \emph{object likelihood functions}, $\like_{\io\iw}(\vcoef_\io) \equiv p(D_{\io\iw}|\vcoef_\io)$, functions only of $\vcoef_\io$ once the data are available.

From the full joint distribution, we can condition on the data to get the joint posterior distribution for the population and SED parameters,
\begin{equation}
p(\vmcoef, \ccoef, \{\vcoef_\io\} | \{D_{\io\iw}\}) = \frac{1}{p(\{D_{\io\iw}\})}
  p(\vmcoef, \ccoef) \prod_{\io=1}^\no \left[ p(\vcoef_\io|\vmcoef, \ccoef)
    \prod_{\iw=1}^{\nw_\io} \like_{\io\iw}(\vcoef_\io) \right],
\label{full-post1}
\end{equation}
where the first factor is the reciprocal of the prior predictive or marginal likelihood, $p(\{D_{\io\iw}\})$, here merely functioning as a normalization constant.
That is, the joint posterior distribution is simply proportional to the joint distribution in equation~\ref{joint1}.

We describe the components in more detail below.

\subsubsection{Nonstationary Gaussian process population model}


As noted above, our goal is to find improved SED estimates for subsequent analysis, not to carefully model the population of SEDs.
We will adopt a tractable model for the \emph{observed} SED population; we do not attempt to account for selection effects that may make the sample not fully representative of the underlying population.%
\footnote{
For our specific downstream purposes---finding a low-dimensional manifold that captures the diversity of galaxy SEDs---selection effects should not significantly compromise discovery of the manifold over the domain accessible to observations.
Put more technically, we will not be using the collection of estimated SEDs to do density estimation on the SED manifold; we will use it only to identify the manifold structure.
For other purposes, selection effects may need to be taken into account; if so, they should be handled in a way that self-consistently handles SED selection effects and measurement errors.
One path forward (not explored here) would be to use the thinned latent marked point process (TLaMPP) framework, previously developed for modeling populations of scalar and vector properties \citep{L04-SourceUncert,LH19-MultilevelHBCosPop-TLaMPP}, generalizing it to handle functional data.
}
We adopt a conceptually simple multivariate normal (Gaussian) population distribution, which still presents computational challenges due to the size of the dataset and the large number of parameters in the model.
The population model adopts a multivariate normal (MVN) distribution for the basis function coefficients, independently for each object, so
\begin{equation}
p(\vcoef_o|\vmcoef, \ccoef) = \Norm(\vcoef_o|\vmcoef, \ccoef)
  \qquad \text{for } o=1 \text{ to } \no,
\label{popn-pdf}
\end{equation}
where $\Norm(\cdot|\cdot,\cdot)$ denotes the probability density function (PDF) for the MVN distribution, $\vmcoef$ is the population mean coefficient vector, and $\ccoef$ is the population covariance describing the dependence between the components of each coefficient vector.

Note that the grouping together of the continuum and line coefficients implies that \emph{$\ccoef$ encodes the mutual dependence between line strength and continuum properties}, for every line.

This $\nb$-dimensional MVN distribution induces finite-rank (``degenerate''), nonstationary Gaussian process (GP) prior and posterior distributions for the population of continuous SEDs produced by the model.
GP regression is becoming widely used in astronomy (see \citet{AF22-GPRegrnAstro} for a review).
GP regression uses a GP to model a \emph{single} function; this requires imposing strong structural assumptions on the GP covariance function (e.g., stationarity).
We use a single GP to model a \emph{population} of functions; the voluminous data enables using a more flexible GP model class.
Further discussion of the relationship between our model, GP regression, and other GP work in astronomy (including the related GP FDA work of \citealt{M+22-HBSED-SNIa}) is in Appendix~\ref{app-gp}.

\subsubsection{Measurement error model}

In this subsection on the measurement error model, we focus on a particular galaxy's SED, and often suppress object indices for clarity.

As noted above, the raw SDSS spectral data are 2D images of cross-dispersed light that get processed to produce the 1D SEDs reported in catalogs.
To quantify uncertainty in a SED's flux density measurements, the SDSS pipeline essentially computes summaries of a Gaussian approximation to the likelihood function for the flux density measured by the data associated with a spectral pixel.

The catalogs report independent errors for each flux density measurement, reflecting an underlying assumption that the raw data contributing to each measurement are statistically independent of the data contributing to other measurements (at least to a good approximation).
Accordingly we consider the data for a SED to have been partitioned into subsets, $D_\iw$, each yielding the flux density estimate at a particular log-wavelength $\lw_\iw$.
Our hierarchical model needs specification of likelihood functions for each flux estimate,
\begin{equation}
\like_\iw(\fden_\iw) \equiv p(D_\iw|\fden_\iw) = C(D_\iw)\, G(\fden_\iw; \fest_\iw, \prcn_\iw),
\label{flux-like}
\end{equation}
where $\fest_\iw$ is the flux estimate for the pixel (e.g., a maximum likelihood estimate), $\prcn_\iw$ is the uncertainty in the estimate quantified as an inverse variance (precision), $G(x; m, \tau)$ denotes a Gaussian function in $x$ (not a PDF for $x$, so it may have an arbitrary normalization), and $C$ is a constant that may depend on the data but not on $\fden_\iw$.
Explicitly,
\begin{equation}
G(\fden_\iw; \fest_\iw, \prcn_\iw) =
  \exp\left[-\frac{1}{2}\prcn_\iw\left(\fden_\iw - \fest_\iw\right)^2\right],
\label{G-def}
\end{equation}
though as a likelihood function, the normalization of $G$ over $\fden_\iw$ is arbitrary, and no harm would be done by using a normalized Gaussian for $G$.
The probability distribution $p(D_\iw|\fden_\iw)$ may be very complicated as a function of the data, $D_\iw$.
But what matters for inference is how it behaves as a function of $\fden_\iw$, and catalogs treat that dependence as well-approximated by a Gaussian, with the Gaussian peak location and width specified by scalar data summaries.

We belabor this description because a number of hierarchical Bayesian analyses in astronomy treat data summaries (quantities analogous to $\fest_\iw$ and $\prcn_\iw$ here) as if they were the data, e.g., using them as nodes in a DAG, or writing the data factors in the model's joint distribution as, say, $p(\fest_\iw|\fden_\iw)$ or $p(\fest_\iw|\fden_\iw, \prcn_\iw)$ (the latter evidently to specify the width of the distribution for $\fest_\iw$), taken to be normal distributions for $\fest_\iw$.
But the flux estimate and its uncertainty are both \emph{complex data products}, i.e., they are both derived from the raw data pertaining to the flux at a particular wavelength: $\fest_\iw = \fest_\iw(D_\iw)$ and $\prcn_\iw = \prcn_\iw(D_\iw)$.
The functions reflect the complex pipeline processing producing these data products.
The distribution $p(\fest_\iw(D_\iw)|\fden_\iw)$ would typically be prohibitively difficult to compute as a distribution for $\fest_\iw$, even though its dependence on $\fden_\iw$---what matters for inference---may be nearly Gaussian.
A distribution of the form $p(\fest_\iw|\fden_\iw, \prcn_\iw)$, used in some studies, is not even a probability for the data as required for HB modeling, because it conditions on a data product, $\prcn_\iw(D_\iw)$.

These formally incorrect ways of using data products as if they were the data are likely motivated by the popularity of probabilistic programming languages for HB modeling, such as \stan\ and \pymc.
Such languages require users to specify probability distributions for nodes in a DAG corresponding to an HB model; they do not permit users to specify likelihood functions, even though that is really all that is needed for terminal data nodes in an HB model.
A way around this is to introduce \emph{surrogate data}, i.e., data whose probability distribution produces a likelihood function with the right dependence on the parameters.
Here, if we think of $\fest_\iw$ a measurement of $\fden_\iw$ with additive Gaussian noise with zero mean and \emph{a prior known} precision $\prcn_\iw$, the sampling distribution PDF for $\fest_\iw$ would be $\Norm(\fest_\iw|\fden_\iw,\prcn_\iw)$, which is proportional to equation~\ref{G-def}.
So treating $\fest_\iw$ as if it were the full data, and $\prcn_\iw$ as if it were specified a priori, yields correct inferences.

\subsubsection{Implementation}

For the SED model, we use $B$-splines for the continuum basis functions, with spline knots chosen so that $\approx 20$ log-wavelength values, $\lw_\iw$, fall between knots, except that we identify a number of regions where there tends to be detailed structure (e.g., sharp edges, or small absorption lines not in the SDSS line list) and use more knots in those regions, with $\approx 10$ log-wavelength values between knots.
For the line profile, $\lp(\cdot)$, we use a Gaussian that has unit area when considered as a PDF over wavelength (not log-wavelength), corresponding to the function used for SDSS line fitting.

Using the population and measurement error components described above, and adopting uniform (constant PDF) priors for $\vmcoef$ and $\ccoef$, the joint posterior for the population and SED parameters, equation~\ref{full-post1}, takes the form
\begin{equation}
p(\vmcoef, \ccoef, \{\vcoef_\io\} | \{D_{\io\iw}\}) \propto 
  \prod_{\io=1}^\no \left[ \Norm(\vcoef_\io | \vmcoef, \ccoef)
    \prod_{\iw=1}^{\nw_\io} G(F_{\io\iw}(\vcoef_\io); \fest_{\io\iw}, \prcn_{\io\iw}) \right],
\label{full-post2}
\end{equation}
with $F_{\io\iw}(\vcoef_\io)$ given by the SED model via equation~\ref{fdens-all}.

To find empirical Bayes SED estimates, we first find the values of the population parameters, $(\hat\vmcoef, \hat\ccoef)$ that maximize the marginal likelihood function for the population parameters,
\begin{equation}
\mlike(\vmcoef, \ccoef) \propto 
  \prod_{\io=1}^\no \int \dif\vcoef_\io\; 
    \left[ \Norm(\vcoef_\io | \vmcoef, \ccoef)
    \prod_{\iw=1}^{\nw_\io} G(F_{\io\iw}(\vcoef_\io); \fest_{\io\iw}, \prcn_{\io\iw}) \right].
\label{mlike-popn}
\end{equation}
Then, \emph{conditional on those maximum marginal likelihood (MML) values}, we compute estimates of the SED coeficients, with uncertainties, which can be used to estimate the SEDs on a grid or to compute functionals of the SEDs.
The MML-conditional $\vcoef_\io$ estimates can be computed analytically; they correspond to weighted least squares (minimum $\chi^2$) estimates, but adjusted by using the estimated population distribution as an informative MVN prior on the coefficients.

The challenging part of this calculation is finding the MML estimates, $(\hat\vmcoef, \hat\ccoef)$.
Maximization of the marginal likelihood function is not possible analytically.
We use an expectation-maximization (EM) algorithm that works with the logarithm of the joint posterior of equation~\ref{full-post2}, in its surrogate-data form:
\begin{equation}
L(\vmcoef, \ccoef, \{\vcoef_\io\}) =
  - \sum_{\io=1}^\no \left[ \log\Norm(\vcoef_\io | \vmcoef, \ccoef)
    \sum_{\iw=1}^{\nw_\io} \log \Norm(\fest_{\io\iw} | \fden_{\io\iw}(\vcoef_\io),\prcn_{\io\iw}) \right].
\label{log-post}
\end{equation}
The logarithms of the multivariate normal PDFs are quadratic forms in $\{\vcoef_\io\}$ and $\vmcoef$.

The EM algorithm begins with an intial choice for $(\vmcoef,\ccoef)$, and then iterates the following two steps:
\begin{enumerate}
\item E-step: Compute $Q(\vmcoef',\ccoef'|\vmcoef,\ccoef) = \expect_{\vcoef} L(\vmcoef', \ccoef', \{\vcoef_\io\})$, where the expectation over all coefficients is done using the current population distribution, i.e., with $\prod_\io \Norm(\vcoef_\io | \vmcoef,\ccoef)$.
\item M-step: Maximize $Q(\vmcoef',\ccoef'|\vmcoef,\ccoef)$ over $(\vmcoef',\ccoef')$ to update the population parameter estimates.
\end{enumerate}
These steps can be computed analytically, thanks to the quadratic forms appearing in the log-marginal-likelihood function; see Appendix~\ref{app-em}.
It can be shown that iterating these steps monotonically increases the marginal likelihood function and converges to a local maximum asymptotically.
Once the MML estimate $(\hat\vmcoef, \hat\ccoef)$ is computed, we can compute estimated coefficients for each SED.

\cnote{Do we use random starts to explore global optimality?}

\cnote{How far did we get with JW's PyTorch optimizer work? Should we mention here that we tried SGD-type optimizers, and EM was superior?}

\section{Application to the SDSS MGS sample}\label{results}

The data modeled here are SED measurements for galaxies comprising the MGS, from SDSS Data Release 17 (DR17; \citealt{SDSS-DR17}), including all available galaxies, excepting those which may have errors in the spectroscopically estimated redshift, amounting to $N = \ngal$ galaxies in total.
We retrieved the MGS using an ADQL query duplicating the selection described by \citet{S+02-SDSS-MGS}.
The spectrum of the $i$th galaxy is measured at $k = 1, \dots, n_i$ wavelengths ($n_i$ ranging from
$2063$ to $3860$), the measurement at each wavelength comprising a
triple $(\tilde \wv_{\io\iw}, \fest_{\io\iw}, \prcn_{\io\iw})$, with lab
frame wavelength $\tilde \wv_{\io\iw}$ in angstroms (\AA), co-added
flux $\fest_{\io\iw}$ in units of $10^{-17}$
erg/s/cm\textsuperscript{2}/\AA, and the precision, or inverse
variance $\prcn_{\io\iw}$ of $f_{i,k}$. 
Each galaxy also has a spectroscopically estimated redshift $z_\io$.
From these measurements, we compute the corresponding rest-frame wavelengths $\wv_{\io\iw} = \frac{\tilde \wv_{\io\iw}}{1+z_\io}$ and take as our data the triples $\{(\wv_{\io\iw}, \fest_{\io\iw}, \prcn_{\io\iw})\}_{\iw=1}^{\nw_\io}$. 
We will also sometimes think of the SEDs as functions of the log-wavel{}ength, denoting $\ll_{\io\iw} = \log_{10} \wv_{\io\iw}$. 
This is convenient because the observations are evenly spaced in log-wavelength, i.e. $\lw_{\io,\iw+1} - \lw_{\io\iw} = \Delta$ does not depend on $\io$ or $\iw$, with the added benefit that (de)redshifting a measured SED becomes a translation operation in terms of the $\lw_{\io\iw}$.
We do not consider redshift uncertainty in our model (it is negligible).

\cnote{Check spec-z errors. Use proper units.}

To model the SEDs for the MGS sample, we used $\nc=153$ B-spline basis functions for the continuum, and $\nl=25$ Gaussian line profiles (corresponding to the number of lines fit by the \texttt{spectro1d} pipeline for the MGS sample), so the total number of basis functions (and coefficients) is $B=178$.
The number of parameters in the SED population model (for the population mean vector and covariance matrix) is $B + B(B+1)/2 = 16{,}109$.
\inote{DK: Please verify the parameter counts.}

\begin{figure}
\includegraphics[width=\textwidth]{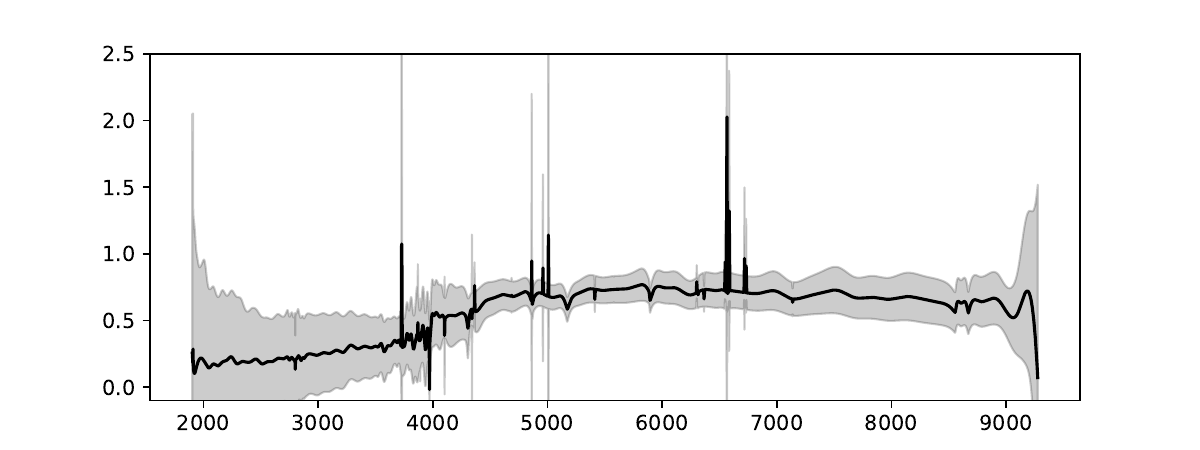}
\caption{The population mean SED as computed by the EM algorithm. 
This is $\fden(\lw) = \sum_{\ib=1}^{\nb}\coef_{0,j}\varphi_j(\ell) + \sum_{j=n_B+1}^{n_B+n_G} \theta_{0,j} \varphi_{j,\cdot}(\ell)$ after fixing line widths so that the line basis functions $\varphi_{j,\cdot}(\ell)$ are well-defined. 
The confidence band is $\pm 1.96 \sigma_\ll$, with $\sigma^2_\ll = \bx^T_\ll\bSigma \bx_\ll$.
\inote{$x$ to be defined}
}
\label{mean_sed}
\end{figure}

\cnote{For DK: The plan from 2023-10-28 email is to switch the line width choice to using the median width for each line.}

Analysis begins by using the EM algorithm to obtain MML estimates of the population parameters, $(\hat\vmcoef, \hat\ccoef)$.
For Figure~\ref{mean_sed}, the coefficients comprising $\hat\vmcoef$ were used in the SED model, equation~\ref{fdmodel-cl}, producing the solid black curve.
The gray band provides a pointwise summary of the dispersion of the SED population about the mean; it spans 1.96 times the pointwise standard deviation (computed using $\hat\ccoef$), producing a 95\% pointwise credible band.

\begin{figure}
\includegraphics[trim={.75cm 0 3cm 0},width=\textwidth]{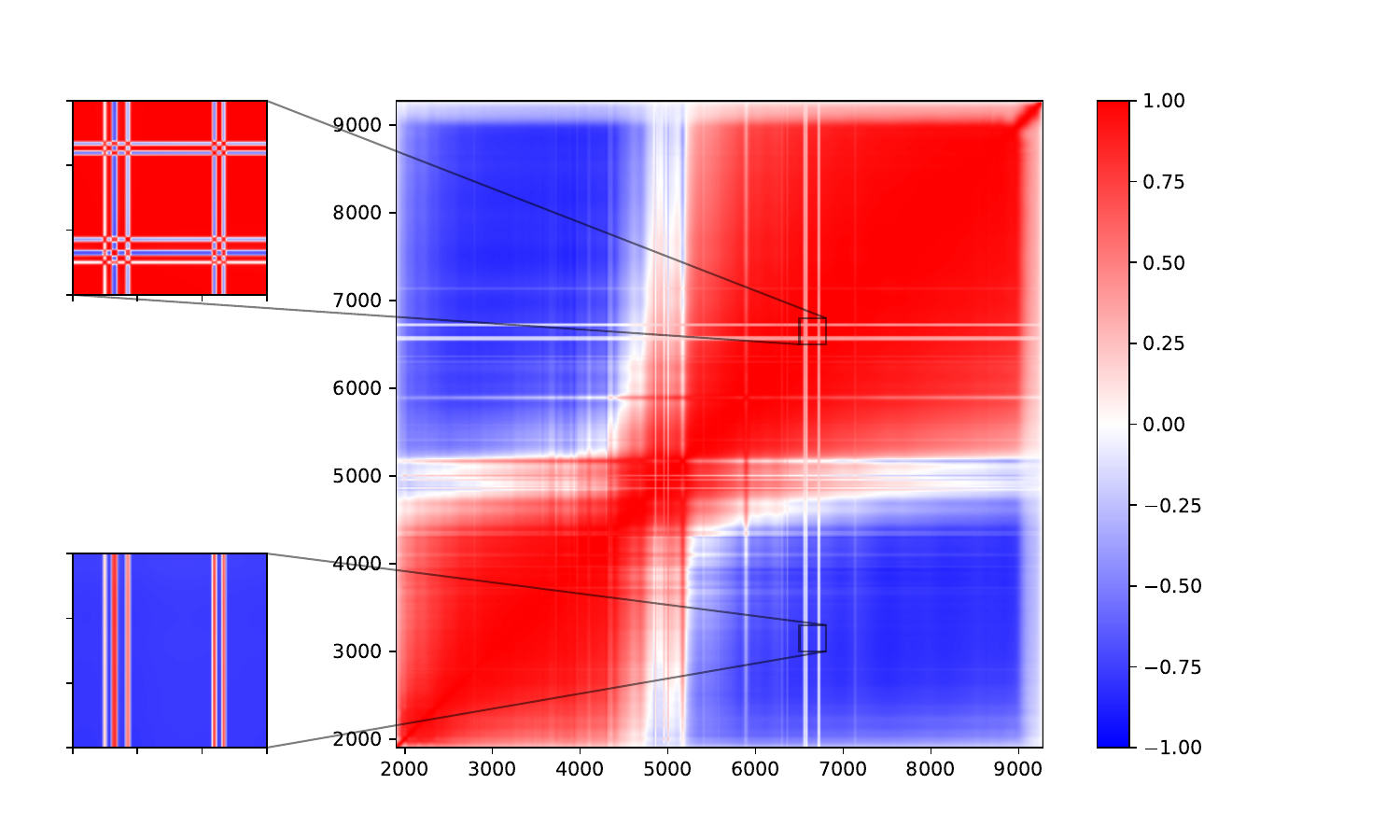}
\caption{With $\fden(\lambda)$ a random SED specified by the estimated population distribution, this plot shows the correlation between SED values at two wavelengths, $C(\lambda_i,\lambda_j) = \Corr(F(\lambda_i),F(\lambda_j))$, as a matrix with values depicted following the colorscale shown on the right (from blue to red, for negative to positive correlations).
Zoomed portions show that the complex of emission lines at 6500--6800 \AA{} are positively correlated with each other, negatively correlated with redder wavelengths and positively correlated with bluer wavelengths.}
\label{corr_mat}
\end{figure}

Figure~\ref{corr_mat} displays the correlations between the values of a SED at two wavelengths, for SEDs drawn from the best-fit population.
The MML value of $\hat\ccoef$ gives the correlations between \emph{coefficients}; these induce correlations between SED values via the linear SED model of equation~\ref{fdmodel-g}.
Two zoomed portions show that the complex of emission lines at 6500--6800 \AA{} are positively correlated with each other, negatively correlated with redder wavelengths and positively correlated with bluer wavelengths.

\cnote{For DK: DR asked for bigger zooms and/or more explanation. TL thinks it's easier to understand the line/continuum correlations from the correlation matrix plot for the coefficients (instead of the flux densities).
TL would like us to show both.
The plan from 2023-10-28 email: Re-create that old figure with current results.}

\cnote{TODO: Highlight the line estimation/prediction capability; it's important for PZ.}

\begin{figure}
\includegraphics[width=\textwidth]{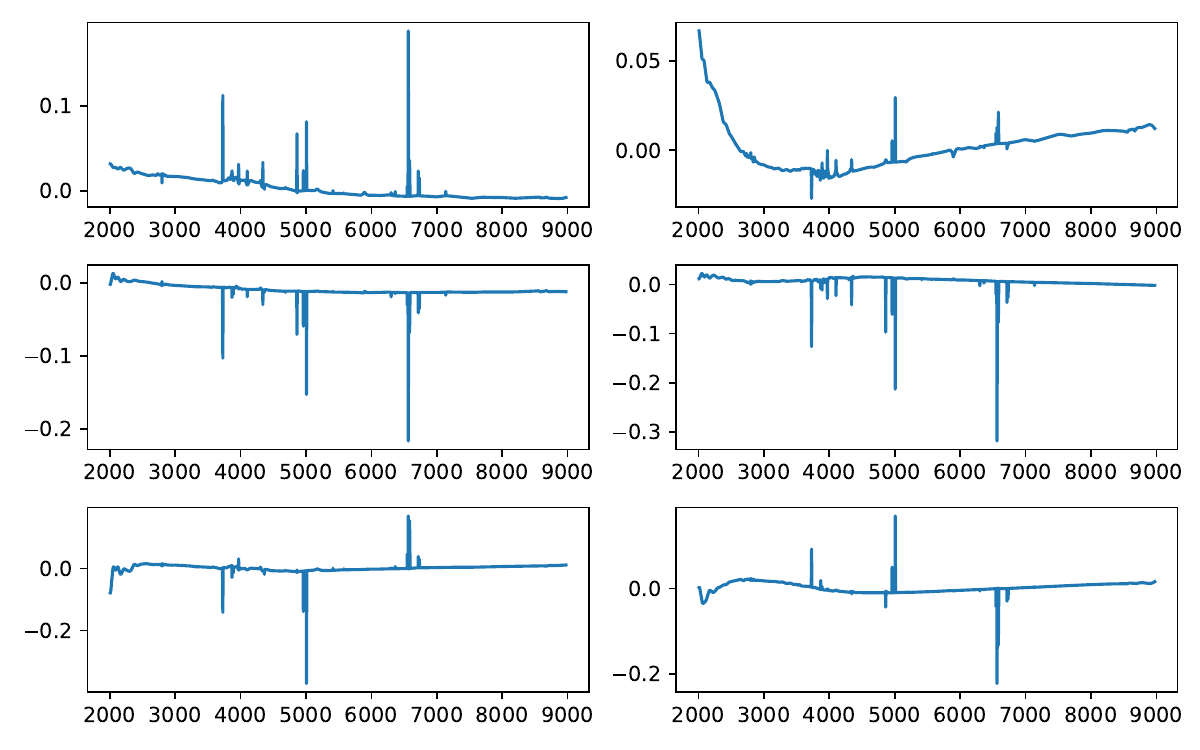}
\caption{The first six eigenfunctions (left--right then top--down) of the covariance function of $\fden(\lambda)$, a random SED with coefficients specified by the estimated population distribution.}
\label{eigens}
\end{figure}

A function sampled from a Gaussian process can be represented as the sum of the population mean function and a weighted sum of eigenfunctions of the GP covariance function, with the weights drawn randomly from independent standard normal distributions scaled by the square roots of the associated eigenvalues.
Figure~\ref{eigens} shows the first six eigenfunctions of the SED covariance function, to give some insight into the dominant structures underlying SED diversity.

\begin{figure}
\includegraphics[width=\textwidth]{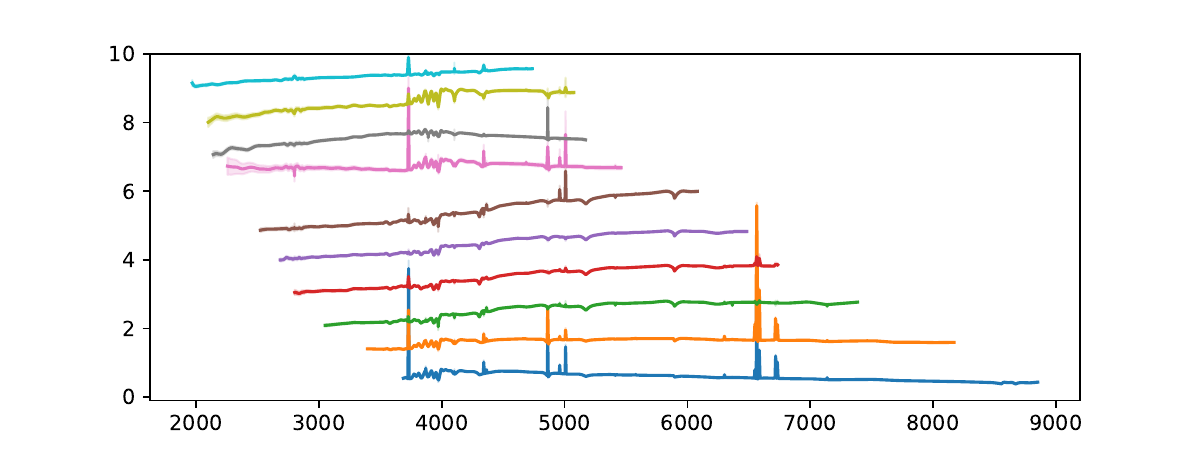}
\caption{Estimates of the 10 SEDs that produced the data shown in Figure~\ref{example_seds}, displayed as a curve showing the estimated SEDs (with colors matching those in the data figure) and an approximate pointwise 95\% confidence band shown as a lighter shade of the curve color.
The confidence bands are quite narrow, except near the short-wavelength end of the SED shown as magenta.}
\label{example_fits}
\end{figure}

The main goal of the model is to allow us to estimate SEDs as continuous functions of wavelength, interpolating between the (unaligned) sample points, $\lw_{\io\iw}$ (and across gaps in the data).
Figure~\ref{example_fits} shows estimates of the SEDs that produced the data shown in Figure~\ref{example_seds}.
That data plot illustrated the challenges arising from rest-frame wavelength grid misalignment, gaps in the data, and noise.
The fitted SEDs are estimated more precisely than a naive look at the spread of the data points might suggest, because the model imposes local smoothness on the estimated SED (via the scales of the basis functions) and enables ``borrowing strength'' across the population, so a SED measured at low signal-to-noise is made to resemble better-measured SEDs that are similar to it.
Note how well the gap in the SED with the green curve is filled in, including the appearance of three small absorption features whose presence is inferred by a kind of ``statistical analogy'' with other SEDs.

\cnote{DK: Perhaps add some transparency to the curves in example fits plot; there are lines that overlap in a confusing way, esp.\ when compared to the last fig.}

\begin{figure}
\includegraphics[width=\textwidth]{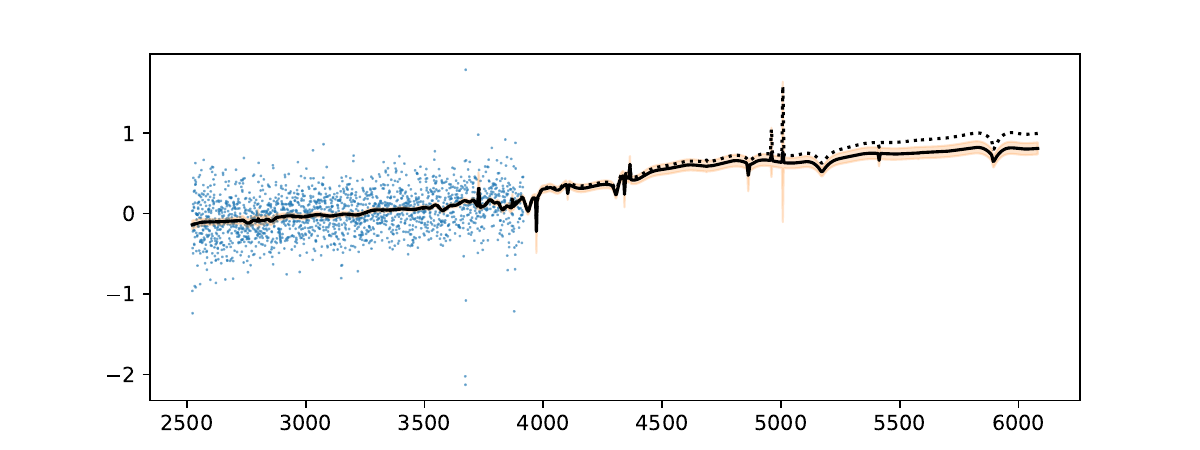}
\includegraphics[width=\textwidth]{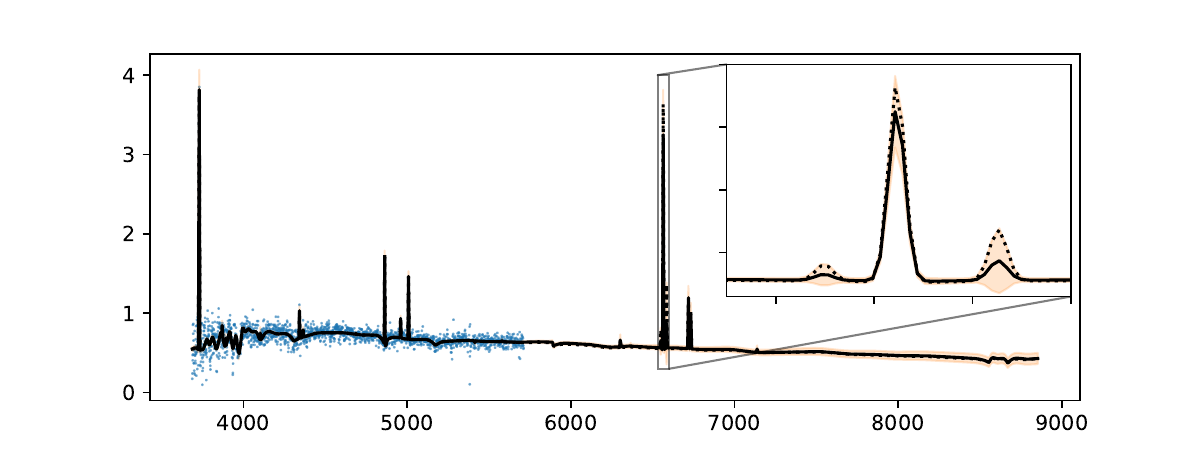}
\caption{Two SEDs from Figures~\ref{example_seds} and \ref{example_fits} (the top here corresponding to the brown case in those figures, the bottom here corresponding to the bottom (dark blue) case), with estimated SED using the full data shown as a dotted line, and estimated SED using only half the data shown as a solid line, with the half of the data used displayed as blue dots.
The pointwise 95\% confidence band corresponds to the fit using only half the data.}
\label{example_extrap1}
\end{figure}

Finally, Figure~\ref{example_extrap1} demonstrates the ability of the model to extrapolate (within the support spanned by the ensemble of measured SEDs).
Two of the 10 SED datasets used in Figures~\ref{example_seds} and \ref{example_fits} are refit here, but omitting half of the data (the full-data fit is also shown for comparison).
The model faithfully recovers the SED even with serious data loss.
Notably, the model is able to recover line features, thanks to how strongly they are correlated with broad continuum properties.

\cnote{For DK, re:  2023-10-28 email, I have swapped the top/bot SEDs in Fig~\ref{example_extrap1}; you suggested coloring them to match the earlier figs.}

\section{Discussion}\label{discuss}

\cnote{TBD; ideas welcome. Probably should mention the limited diversity of MGS. 
Something I've wondered about: if we include LRGs and/or QSOs, perhaps we should model each population separately.
Are there good diagnostics that could guide decisions about this?}

Conceptually, the SnL model appears deceptively simple: It adopts a linear model for SEDs, and a multivariate Gaussian population distribution for the coefficients in the SED model.
So it may seem surprising that it has the capabilities demonstrated in the previous section, in particular, the ability to fill in large parts of a SED with missing data, including extrapolation.
But the linear/Gaussian description hides nontrivial complexity:  the population distribution has $\sim 10^4$ parameters and, via the linear flux density model, corresponds to a nonstationary Gaussian process SED model that can learn correlations of every region of the analyzed SEDs with every other region.
In particular, the line coefficients for a SED (corresponding to the flux in each line) are allowed to depend on the entire SED (including the strengths of other lines).
This flexibility is possible because the dataset is large, comprising over a billion measurements.
A gap can be filled because there is a lot of information from other SEDs in that region of the spectrum, shareable because of the gridless basis function representation.
A modest amount of extrapolation is possible with good fidelity for many SEDs because the diversity of redshifts across the sample makes information in the extrapolation region available from other SEDs.
The linearity and Gaussianity enable analytical implementation of many steps of the analysis (in particular, the EM algorithm), so despite its implicit complexity, the model is scalable to large datasets, which is the key to learning the many parameters that give SnL its flexibility.

The SDSS MGS comprises mainly regular galaxies, though it does include some active galactic nuclei (AGN) that are not classified as quasi-stellar objects (QSOs)---e.g., it includes some Seyfert galaxies---and it includes a subset of the SDSS legacy survey luminous red galaxy (LRG) sample.
An interesting question is whether the SnL model is flexible enough to handle a more diverse collection of SEDs, e.g., including the legacy QSO and LRG samples.
We have focused here on the MGS because it is widely studied, particularly for testing photo-$z$ methods.
We leave for future work exploring whether a more diverse galaxy sample can be accommodated by SnL, or whether separate analyses are needed for other populations.

\begin{acknowledgments}
This material is based upon work supported by the National Science Foundation under Grant No. AST-1814840 (Cornell University) and Grant No. AST-1814778 (Johns Hopkins University).
\end{acknowledgments}

\bibliography{SnLRefs.bib}

\appendix

\section{Connection to Gaussian processes}\label{app-gp}

A stochastic process is a rule for generating joint distributions for the values of a function at an arbitrary set of sample points, with the rule ensuring that the marginal distributions for different finite sets of sample points are mutually consistent.
A GP is a stochastic process defined so that the joint distribution for every finite set of function values is a MVN distribution.
The mutual consistency requirement is that the MVN distribution for a set of function values at points that are a subset of a larger set must correspond to the MVN distribution for that larger set of function values, marginalized over the values at the omitted points.

One way to construct a GP is to specify a mean function, $\mu(\lambda)$, and a covariance function, $c(\lambda,\lambda')$.
For a function of wavelength measured at sample points $\lambda_k$, the induced MVN has a mean vector  with components $\mu(\lambda_k)$ , and a covariance matrix with components $c(\lambda_k,\lambda_{k'})$.
This construction gives the resulting family of distributions the marginalization consistency properties required of a valid stochastic process.
Alternatively, a GP may be constructed by putting a MVN distribution on the coefficients of a basis function representation of the target function.
The two constructions are related.
The covariance function construction corresponds to a basis expansion using eigenfunctions of the covariance function.
The basis function construction corresponds to using a covariance function computed using pairwise products of basis functions.
See \citet{RW06-GP4ML} for details.

In GP regression, a GP prior is used for nonparametric curve fitting to a dataset comprising samples (point evaluations or functionals) of a \emph{single} function, say, a spectrum (function of wavelength) or a light curve (function of time).
If the samples are noiseless or have additive Gaussian noise, Bayesian fitting of the samples produces a posterior GP for the sampled function with an updated mean and covariance function.
Focusing on the function values at the $N$ sample points, GP regression corresponds to estimating an $N$-dimensional mean function and a symmetric $N\times N$ covariance matrix with $N(N+1)/2$ unique components.
There are many more potential unknowns than there are sample points.
Useful inference is only possible by imposing structure that reduces the degrees of freedom.
Common structural assumptions include using a constant mean function (with a single scalar parameter), and a stationary covariance function, $c(x,x') = k(x-x';\eta)$, with a parameter vector $\eta$ with just two or three parameters, so the $N(N+1)/2$ unique entries in the covariance matrix are determined by just those few parameters.

We are instead using a GP for FDA (see, e.g., \citealt{SCQ11-GPFDA}), i.e., for describing a \emph{collection} of related measured functions.
If a single function has $N$ sample points, and we observe $M$ functions, we have $N\times M$ total observations.
When $M$ is large, this can be enough to tightly constrain the full $N\times N$ covariance matrix.
This is the case here, where $N\sim 10^3$, $M\sim 10^6$, and we use a few hundred basis functions, so that the covariance matrix has $\sim 10^4$ parameters.
(For SED fitting, the sample points differ in number and location across the $M$ measured functions, but this is addressed by representing the measured functions in terms of shared basis functions.)
Explicitly, the covariance function in our model is $c(\lw,\lw') = \vgb^T(\lw)\cdot\ccoef\cdot\vgb(\lw')$, and thus is determined by the $B\times B$ covariance matrix for the basis function coefficients, $\ccoef$, where in our application, $B\sim 10^2$.

\citet{M+22-HBSED-SNIa} adopt a qualitatively similar construction for modeling a collection of Type~Ia supernova (SN~Ia) SEDs, including time evolution.
In their case, both the SED sampling (corresponding to $N$) and the number of SEDs (corresponding to $M$) are much smaller than in our case.
This both enables and motivates a fully hierarchical Bayesian treatment (i.e., using MCMC to explore the population parameter space), since parameter uncertainties are significant.
In our application, the data are much more voluminous; a fully Bayesian treatment would be computationally expensive, but an empirical Bayes approximation---optimizing over population parameters, and analytically marginalizing over other parameters---is feasible and accurate.

\cnote{That reflects TL's current understanding from a quick look at the Mandel paper; we should have a closer look later and revise if needed.}

\section{EM algorithm}\label{app-em}

\cnote{DK: Please double-check my translation. It was done by copy-and-pasting your equations and tediously working through them, so it should be pretty close to right.}

As a recap from \S~\ref{model}, we estimate the SED population parameters, $(\vmcoef, \ccoef)$, using an expectation-maximization (EM) algorithm that works with the logarithm of the joint posterior of equation~\ref{full-post2}, in its surrogate-data form:
\begin{equation}
L(\vmcoef, \ccoef, \{\vcoef_\io\}) =
  - \sum_{\io=1}^\no \left[ \log\Norm(\vcoef_\io | \vmcoef, \ccoef)
    \sum_{\iw=1}^{\nw_\io} \log \Norm(\fest_{\io\iw} | \fden_{\io\iw}(\vcoef_\io),\prcn_{\io\iw}) \right].
\label{log-post-em}
\end{equation}
This is called the \emph{complete-data log-likelihood} in the EM literature, which views problems like this as ``missing data problems,'' with ``missing data'' referring, not solely to observables that were lost or inaccessible, but also to parameters that, if known, would simplify computation of the likelihood; here the missing data are the coefficients, $\{\vcoef_\io\}$.

The EM algorithm begins with an intial choice for $(\vmcoef,\ccoef)$, and then iterates the following two steps:
\begin{enumerate}
\item E-step: Compute $Q(\vmcoef',\ccoef'|\vmcoef,\ccoef) = \expect_{\vcoef} L(\vmcoef', \ccoef', \{\vcoef_\io\})$, where the expectation over all coefficients is done using the current population distribution, i.e., with $\prod_\io \Norm(\vcoef_\io | \vmcoef,\ccoef)$.
\item M-step: Maximize $Q(\vmcoef',\ccoef'|\vmcoef,\ccoef)$ over $(\vmcoef',\ccoef')$ to update the population parameter estimates.
\end{enumerate}
The logarithms of the multivariate normal PDFs in equation~\ref{log-post-em} contain quadratic forms in $\{\vcoef_\io\}$ and $\vmcoef$, and, from the normalization constants, logarithms of the determinant of $\ccoef$.
As a result, the quantities in both the E-step and M-step can be computed analytically.

To compute the results for the E-step and M-step we need notation related to values of the basis functions on the data's SED sample points (in wavelength).
Let $\dmat_\io$ be the $\nw_\io\times\nb$ \emph{design matrix for object $\io$} collecting the basis functions evaluated at the log-wavelength points associated with the flux estimates for that object; its $(\iw,\ib)$th entry is $\gb_b(\lw_{\io\iw})$.
Let $\pmat_\io$ be the \emph{precision matrix for object $\io$}, a diagonal matrix collecting the precision values for the flux measurements for object $\io$; its $(\iw,iw)$th entry is $\prcn_{\io\iw}$.
From the surrogate data perspective (corresponding to the final normal distribution terms in equation~\ref{log-post-em}), the SED model's prediction for an object's flux estimates is $\expect\left[\vfest_\io\right] = \dmat_\io \vcoef_\io$, and the covariance matrix for the flux estimates is $\cov\left[\vfest_\io\right] = \pmat^{-1}_\io$.

\subsection{The E-step}

The complete data log-likelihood is, absorbing everything not involving $\vmcoef$ or $\ccoef$ into the constant $c$:
\begin{equation}
L(\vmcoef, \ccoef, \{\vcoef_\io\}) =
  c - \sum_{\io=1}^\no + \frac12 \log |\bSigma| + \frac12
    \vcoef_\io^T\bSigma^{-1}\vcoef_\io - \vcoef_\io^T \ccoef^{-1} \vmcoef 
    + \frac12 \vmcoef^T \bSigma^{-1}\vmcoef.
\end{equation}
The E-step computes the expectation of this with respect to the population distribution for $\vcoef_\io$, using the current population parameters.
Let
\begin{equation}
\bC_\io = (\ccoef^{-1} + \dmat_\io^T \pmat_\io \dmat_\io)^{-1}.
\end{equation}
Then the conditional mean coefficient vector for object $\io$ is
\begin{equation}
\vcmcoef_\io 
  \equiv \expect\left[\vcoef_\io | \vfest_\io,\vmcoef,\ccoef\right] 
  = \bC_\io \dmat_\io^T \dmat_\io \vfest_\io + \bC_\io \ccoef^{-1} \vmcoef.
\end{equation}
Using this, the expectation value of the coefficient-dependent quadratic form in the log-likelihood is
\begin{equation}
\expect\left[\vcoef_\io^T(\ccoef')^{-1}\vcoef_\io | \vfest_\io,\vmcoef,\ccoef\right]
  = \tr\left[(\ccoef')^{-1}\bC_\io\right] + \vcmcoef_\io (\ccoef')^{-1} \vcmcoef_\io.
\end{equation}
With these results we can compute the result of the E-step: the objective function, $Q$,
\begin{equation}
\begin{aligned}
Q(\vmcoef',\ccoef'|\vmcoef,\ccoef) &=
  - \sum_{\io=1}^\no \left\{ \frac12 \log |\ccoef'| 
    + \frac12 \tr\left[(\ccoef')^{-1}\bC_\io\right] 
    + \frac12 \vcmcoef_\io (\ccoef')^{-1} \vcmcoef_\io \right.\\
    &\qquad \left. - (\vcmcoef_\io^T (\ccoef')^{-1} \btheta_0 
    + \frac12 (\vmcoef')^T (\ccoef')^{-1}\vmcoef' \right\}.
\end{aligned}
\end{equation}
Note that the $(\vmcoef,\ccoef)$ dependence enters via $\vcmcoef_\io$ and $\bC_\io$.

\subsection{The M-step}

Now we must find the population parameter estimates to use for the next iteration,
\begin{equation}
\hat\vmcoef', \hat\ccoef' = 
  \argmax_{\vmcoef',\ccoef'} Q(\vmcoef',\ccoef'|\vmcoef,\ccoef),
\end{equation}
which we can do in the usual way, requiring partial derivatives to vanish.

First, we have that
\begin{equation}
\frac{\partial Q}{\partial \vmcoef'} = 
  \sum_{\io=1}^\no \left[2(\ccoef')^{-1}\vcmcoef_\io 
    - 2 (\ccoef')^{-1}\vmcoef'\right].
\end{equation}
Requiring this to vanish and solving, we see that $Q$ is maximized at $\hat\vmcoef' = \frac1N \sum_{i=1}^N \vcmcoef_\io$ regardless of $\ccoef'$.

Finding $\hat\ccoef'$ requires computing a number of derivatives with respect to the matrix $\ccoef'$.
After some nontrivial linear algebra we find the fairly simple result,
\begin{equation}
\hat\ccoef' 
  = \frac1N \sum_{\io=1}^\no \bC_\io +\vcmcoef_\io(\vcmcoef_\io)^T - \hat\vmcoef'(\hat\vmcoef')^T.
\end{equation}

\end{document}